\documentclass[aps,prd,reprint,superscriptaddress,nofootinbib,floatfix]{revtex4-2}

\usepackage{graphicx}
\usepackage{amsmath}
\usepackage{mathrsfs}

\usepackage[breaklinks=true,colorlinks=true,
            linkcolor=blue,citecolor=blue,urlcolor=blue]{hyperref}
\AtBeginDocument{\let\linenumbers\relax}

\newcommand{\ud}{\mathrm{d}}

\providecommand{\apjl}{Astrophys.\ J.\ Lett.}
\providecommand{\aap}{Astron.\ Astrophys.}

\begin{document}

\title{Identification of Lensed Gravitational-Wave Beat Patterns by LISA}

\author{Hengyu Wu}
\affiliation{School of Physics and Optoelectronic Engineering, Yangtze University, Jingzhou 434023, China}
\author{Tonghua Liu}
\email{liutongh@yangtzeu.edu.cn}
\affiliation{School of Physics and Optoelectronic Engineering, Yangtze University, Jingzhou 434023, China}
\author{Kai Liao}
\email{liaokai@whu.edu.cn}
\affiliation{School of Physics and Technology, Wuhan University, Wuhan 430072, China}

\date{\today}

\begin{abstract}
Strong lensing of massive black hole binaries can produce multiple gravitational-wave images with different magnifications and arrival times. LISA signals remain in band for months to years, allowing multiple lensed images to overlap during the inspiral stage and generate beat patterns. A singular isothermal sphere lens model is adopted to describe the lensing configuration, and two-image beat waveforms are constructed from massive black hole binary signals. To isolate the beat pattern itself, waveform mismatch is evaluated only during the overlapping inspiral stage before the coalescence of the first image, excluding contributions from the delayed merger peak of the second image. Using the HS-nod-SN (B+20) strong-lensing population, the occurrence rate of identifiable beat events is estimated, and Bayesian parameter estimation is performed with a beat template. Beat patterns are most readily identified when the lensing time delay is short and the delayed image has a relatively large magnification. Among 196 detectable two-image lensed events, 92 satisfy the temporal-overlap condition and 14 satisfy the beat-identification criterion, corresponding to an identifiable beat fraction of about 7\%. Posterior inference shows that the beat template can recover the lensing time delay and magnification parameters for a representative beat event. These results indicate that lensed beat patterns constitute a distinguishable subset of strongly lensed LISA events and provide a unique observational signature of strong lensing in the LISA band.
\end{abstract}

\maketitle

\section{Introduction} 
\label{sec-int}

Gravitational waves (GWs) were first directly detected in 2016 by the LIGO-Virgo collaboration \citep{2016PhRvL.116f1102A}, experimentally verifying the gravitational wave prediction of general relativity originally proposed by Einstein in 1916 and 1918 \citep{1916SPAW.......688E,1918SPAW.......154E}. This landmark discovery opened a new observational window into the Universe and inaugurated the era of gravitational wave astronomy and multimessenger astrophysics. Subsequently, GW170817 marked the first direct observation of a binary neutron star merger \citep{2018PhRvL.120c1104A}, while GW200105 and GW200115 represented the first credible detections of black hole--neutron star coalescence events \citep{2021ApJ...915L...5A}.

When electromagnetic or gravitational waves emitted from distant astrophysical sources propagate across cosmological distances toward the observer, their trajectories are deflected by intervening gravitational potentials, leading to distortions of the spacetime geometry along the propagation path \citep{1936Sci....84..506E,1971PhRvD...3.3239L,2013RAA....13...15C,2014MNRAS.443..969C,1992grle.book.....S,2010ARA&A..48...87T}. This phenomenon is known as gravitational lensing. Theoretical descriptions of gravitational lensing are generally formulated within two complementary regimes: geometric optics and wave optics. In the geometric-optic limit, lensing effects are characterized by multiple images, magnifications, and time delays, and this framework successfully describes most currently observed lensing phenomena \citep{1964MNRAS.128..307R,2017ApJ...835..103T}. In contrast, within the wave-optics regime, diffraction and interference effects become important, leading to frequency-dependent modulations of the signal amplitude and phase, including phenomena analogous to the Poisson-Arago spot and diffraction fringes \citep{1999PThPS.133..137N,1998PhRvL..80.1138N,2003ApJ...595.1039T,2019ApJ...875..139L,2022JCAP...07..022B,2021SCPMA..6420462Z}.

To date, gravitational lensing has been firmly established only through electromagnetic observations. However, with the rapidly increasing number of detected gravitational-wave events \citep{2025arXiv250818082T,2026arXiv260527223T}, lensed gravitational waves have emerged as an important topic in modern astrophysics and cosmology \citep{2026PhRvD.113h3009L}. Compared with electromagnetic signals, lensed gravitational waves possess several distinctive features. First, the wavelength of gravitational waves is typically many orders of magnitude larger than that of electromagnetic radiation \citep{1983ApJ...271..551O,1975Ap&SS..34L...7B,1981Ap&SS..78..199B}, making wave effects potentially significant when the characteristic lens scale becomes comparable to the gravitational-wave wavelength. Second, gravitational waves produced during the inspiral phase of compact binary coalescences exhibit a high degree of phase coherence and can often be approximated as quasi-monochromatic signals over sufficiently short timescales. Such coherence makes the superposition of multiple lensed gravitational-wave images particularly favorable for generating observable interference phenomena \citep{1981SvAL....7..213M,1985A&A...148..369S,1985BAAS...17..907D,1991ApJ...374L...5P,2005PhRvD..71j1301Y}.

When the source, lens, and observer are sufficiently aligned, the system enters the strong-lensing regime, producing multiple magnified images associated with different propagation paths and corresponding lensing time delay $\Delta t_{\rm lens}$. For gravitational waves, signals propagating along different lensing trajectories similarly acquire distinct arrival times. Under suitable conditions, multiple lensed gravitational-wave images may overlap simultaneously within the detector observation window, giving rise to interference effects. During the inspiral stage of compact binaries, the gravitational-wave frequency evolves slowly with time, such that different lensed images generally differ only by a small instantaneous frequency shift $\Delta f$. When $\Delta f$ is sufficiently small, the coherent superposition of these signals naturally generates a characteristic beat pattern \citep{2021MNRAS.507..761H,2020PhRvD.101f4011H}. Such beat signals may not only encode information about the lensing system itself, but also provide a potential probe of cosmological parameters and lens properties.

Gravitational lensing possesses broad applications in astrophysics and cosmology, including dark matter probing \citep{2009PhRvD..80j4009C,2013PhRvL.110o1103C,2019PhRvD..99h3526C,2019PhRvL.122d1103J}, constraints on the propagation speed of gravity \citep{2017PhRvL.118i1102F,2017PhRvL.118i1101C}, cosmological parameter estimation \citep{2010PhRvL.105y1101S,2011MNRAS.415.2773S,2017NatCo...8.1148L,2024ApJ...965L..11L}, and tests of the wave nature of gravitational waves. Although no conclusive lensed gravitational-wave events have been detected so far, the next generation of high-sensitivity detectors, such as Laser Interferometer Space Antenna (LISA) \citep{2017arXiv170200786A,2019BAAS...51g..77T,2024arXiv240207571C,2025PhRvD.112l3512G,2023PhRvD.108l3543C}, is expected to significantly enhance the possibility of observing such signals in the near future. At present, cosmology is facing the severe Hubble tension problem, where two major observational approaches yield significantly different values of the Hubble constant, motivating the need for new and independent cosmological probes \citep{2019ApJ...876...85R,2021A&A...652C...4P,2021CQGra..38o3001D}. As standard sirens, gravitational waves can bypass the systematic uncertainties associated with the traditional cosmic distance ladder, directly measure luminosity distances, and, when combined with redshift information, provide independent constraints on cosmological parameters. In particular, gravitational wave beat signals generated by lensing-induced interference not only exhibit higher waveform distinguishability but also retain the essential advantages of lensed gravitational waves for fundamental physics and cosmological studies. Consequently, such beat signals can likewise serve as powerful probes for dark matter studies, cosmological parameter measurements, tests of gravity, and investigations of gravitational-wave propagation, providing a promising new observational window for future cosmology and fundamental physics.

For current ground-based gravitational-wave detectors, including the LIGO--Virgo--KAGRA (LVK) network \citep{2015CQGra..32g4001L,2015CQGra..32b4001A,2023PhRvX..13d1039A,2012CQGra..29l4007S,2013PhRvD..88d3007A,2021PTEP.2021eA101A}, as well as future third-generation observatories such as the Einstein Telescope (ET) \citep{2020JCAP...03..050M,2026JCAP...03..081A} and Cosmic Explorer (CE) \citep{2019BAAS...51g..35R,2021arXiv210909882E}, the primary targets are high-frequency gravitational-wave signals emitted during the late inspiral and merger stages of compact binaries. These signals typically remain in band only for milliseconds to hundreds of seconds. In contrast, the space-based detector LISA operates in the low-frequency band ($10^{-4}\text{--}0.1~\mathrm{Hz}$), corresponding to much longer gravitational-wave wavelengths. Significant geometric-optics lensing in this regime generally requires lens masses of at least $\sim10^{8}M_{\odot}$ \citep{2004GReGr..36..983V,2020MNRAS.492.1127M}. For galaxy-scale strong lenses, the associated time delays are typically on the order of days to months \citep{2024SSRv..220...19N}. Consequently, for ground-based detectors, the duration of observed gravitational-wave signals is usually far shorter than the lensing time delay, preventing different lensed images from overlapping coherently within the detector time window. By contrast, low-frequency gravitational-wave signals observed by LISA may persist for months to years, substantially increasing the probability that multiple lensed images overlap simultaneously in the detector data stream and generate observable interference and beat phenomena. Motivated by this unique capability, in this work we focus on the beat signatures produced by strongly lensed gravitational waves in the LISA band.

This work investigates strongly lensed gravitational-wave beat patterns produced by overlapping images of massive black hole binaries in the LISA band. The lensing configuration is described within the singular isothermal sphere (SIS) framework \citep{1992ARA&A..30..311B,1992grle.book.....S}, and the formation of beat patterns arising from the coherent superposition of multiple lensed images is analyzed. The resulting signals are characterized in both the time and frequency domains, and a mismatch-based criterion is introduced to determine when the beat modulation becomes observationally distinguishable. Based on recent predictions for the population of strongly lensed LISA sources, the expected occurrence rate of identifiable beat events is estimated. Bayesian inference is further applied to simulated beat signals to assess the recoverability of the lensing magnifications and time-delay parameters. The structure of the paper is as follows. Sect. \ref{sec-glgw} introduces the lensing framework and event-rate model, Sect. \ref{sec-iog} discusses the formation and properties of beat patterns, Sect. \ref{sec-methods} presents the identification and parameter-estimation methodology, Sect. \ref{sec-results} reports the main results, and Sect. \ref{sec-discussion-conclusion} summarizes the conclusions.

\section{Strong Lensing of Gravitational Waves} 
\label{sec-glgw}

To investigate the impact of strong lensing on beat signals observed by LISA, we adopt the SIS model, which is widely used to describe galaxy-scale strong lenses and provides a simple framework for characterizing the formation of multiple images.

\subsection{SIS Model}

To describe the gravitational lensing effect produced by galaxy-scale structures, various lens models have been developed in the literature. Among them, the SIS model is one of the most commonly adopted approximations due to its simplicity and its ability to capture the essential properties of massive early-type galaxies \citep{1992grle.book.....S,2006EAS....20..161K,2006ApJ...649..599K}. In this model, the lens is assumed to possess a spherically symmetric mass distribution with density profile
\begin{equation}
\rho(r)\propto r^{-2},
\end{equation}
which corresponds to an isothermal system supported by a constant one-dimensional velocity dispersion $\sigma_v$. Since elliptical galaxies contribute dominantly to the strong-lensing optical depth, the SIS profile provides a useful and sufficiently accurate framework for studying strongly lensed gravitational-wave events.

For a source located at an angular position $\beta$ relative to the optical axis, the lens equation of the SIS model yields two possible image positions,
\begin{equation}
\theta_{\pm}=\beta\pm\theta_E,
\end{equation}
where $\theta_{\pm}$ denote the angular locations of the two lensed images. The quantity
\begin{equation}
\theta_E =4\pi\left(\frac{\sigma_v}{c}\right)^2\frac{D_{LS}}{D_S},
\end{equation}
is the Einstein radius, which characterizes the angular scale of the lensing configuration. Here, $D_L$, $D_S$, and $D_{LS}$ represent the angular diameter distances between the observer and the lens, the observer and the source, and the lens and the source, respectively.

For convenience, one usually introduces the dimensionless source position
\begin{equation}
y=\frac{\beta}{\theta_E}.
\end{equation}
When the source lies inside the Einstein radius, namely $y<1$, two distinct lensed images are produced. Otherwise, only a single image can be observed. In the geometric optics regime relevant for most LISA massive black hole binary signals considered in this work, these multiple images correspond to repeated gravitational-wave signals arriving at the detector with different amplitudes and arrival times.

Assuming a spatially flat Friedmann Robertson Walker cosmology, the angular diameter distance to redshift $z$ is given by
\begin{equation}
D_A(z)=\frac{1}{H_0(1+z)}
\int_0^z\frac{dz'}{E(z')},
\end{equation}
where
\begin{equation}
E(z)=\left[
\Omega_M(1+z)^3+\Omega_\Lambda
\right]^{1/2}.
\end{equation}
Here, $\Omega_M$ and $\Omega_\Lambda$ denote the present matter and dark-energy density parameters, respectively. The radiation component is neglected because its contribution at low redshift is negligible. Throughout this work, a flat $\Lambda$CDM cosmology with $H_0 = 67.66~{\rm km\,s^{-1}\,Mpc^{-1}}$ and $\Omega_M = 0.31$ is adopted, consistent with the Planck 2018 cosmological parameters \citep{2020A&A...641A...6P}.

The two lensed gravitational-wave signals propagate along different effective paths in spacetime and therefore reach the detector at different times. For the SIS lens, the time delay between the two images can be expressed as
\begin{equation}
\Delta t_{\rm lens} = y\,\Delta t_z,
\end{equation}
with
\begin{equation}
\Delta t_z=
32\pi^2
\left(\frac{\sigma_v}{c}\right)^4
(1+z_L)
\frac{D_LD_{LS}}{cD_S},
\end{equation}
where $z_L$ is the redshift of the lens. For galaxy-scale strong lensing systems, the corresponding delay is typically on the order of days to months, which is particularly important for space-based gravitational-wave detectors due to the long signal duration of inspiraling massive black hole binaries.

In addition to the time delay, lensing also modifies the observed amplitudes of the two images through the amplitude magnification factors
\begin{equation}
\mu_{\pm}=
\sqrt{\frac{1}{y}\pm1},
\label{eq:sis_mag}
\end{equation}
where $\mu_\pm$ denote the amplitude magnification factors entering the waveform model, corresponding to the square roots of the conventional SIS lensing magnifications.

As a result, the corresponding signal-to-noise ratios (SNRs) \citep{2013LRR....16....9Y} become
\begin{equation}
\rho_{\pm}=\mu_{\pm}\rho_0,
\end{equation}
where $\rho_0$ is the intrinsic unlensed SNR. Consequently, one image is amplified more strongly while the other is relatively weaker. In realistic observations, both images must generally exceed the detector sensitivity threshold in order to be identified as a strongly lensed gravitational-wave event.

\subsection{LISA strong-lensing event-rate model}
\label{sec-lensing-rate-model}

The observational relevance of lensed beat signals depends on the expected strong-lensing rate of massive black hole binaries in LISA. We adopt the recent LISA strong-lensing rate estimates of \citet{2025PhRvD.112l3512G}, which considered several massive black hole binary source populations and different galaxy-lens evolution models. In this work, we use the heavy-seed no-delay model with supernova feedback, HS-nod-SN (B+20), as the representative source population. This model provides a useful parent sample for our purpose because most of its strongly lensed detections have two observable images, which is the configuration required for an overlapping beat signal.

\begin{table}[!t]
\caption{Expected number of detectable strongly lensed events for the HS-nod-SN (B+20) source population in a 4-year LISA mission, based on Ref.~\cite{2025PhRvD.112l3512G}.}
\label{tab-lisa-rate}
\centering
\renewcommand{\arraystretch}{1.4}
\begin{tabular}{lccc}
\hline\hline
Lens population & 1 image & 2 images & Total \\
\hline
$z$-independent & 18 & 196 & 214 \\
$z$-dependent 1 & 11 & 121 & 132 \\
$z$-dependent 2 & 14 & 154 & 168 \\
\hline
\end{tabular}
\end{table}

Table~\ref{tab-lisa-rate} summarizes the predicted numbers of detectable strongly lensed events reported by \citet{2025PhRvD.112l3512G} for a four-year LISA mission. Since beat formation requires the overlap of two lensed images in the LISA band, the two-image subset is the most relevant sample for this study. In the following analysis, the 196 detectable two-image lensed events are used as the parent population for estimating the fraction of identifiable beat events.
\section{Beat Patterns in Lensed LISA Signals}
\label{sec-iog}

\begin{figure}[!t]
  \centering
  \includegraphics[width=0.95\columnwidth]{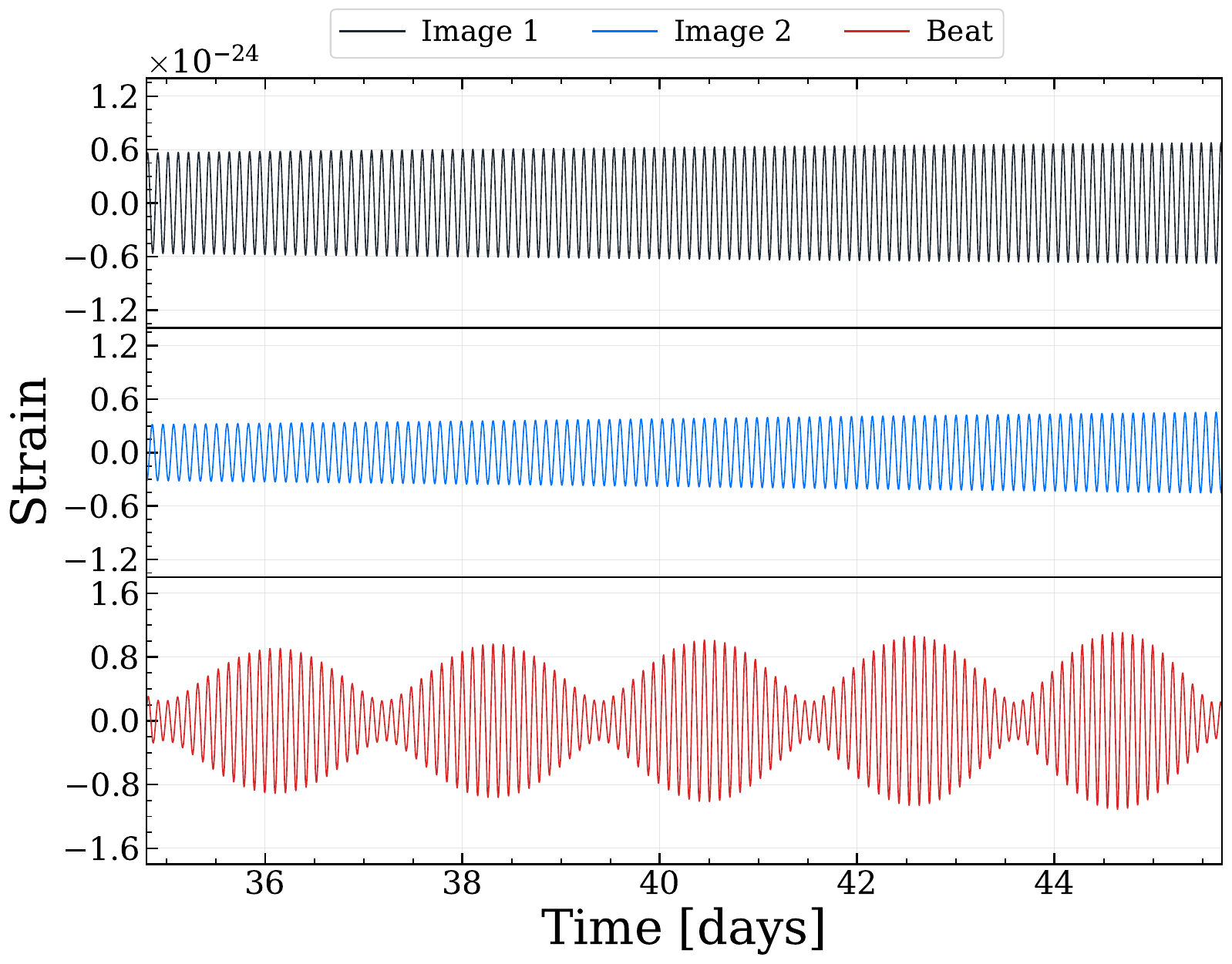}
  \caption{Time-domain illustration of the formation of a lensed gravitational-wave beat pattern. The top and middle panels show the two lensed images in the overlapping inspiral stage, while the bottom panel shows their coherent superposition. The slowly varying amplitude envelope in the bottom panel is the beat modulation produced by the interference of the two lensed images.}
  \label{fig-int}
\end{figure}

\subsection{Time-domain formation of beat patterns}
\label{sec-beat-g}

When the two lensed GW signals arrive at the detector simultaneously during a certain observation interval, the total strain can be written as $h(t)=h_1(t)+h_2(t)$ \citep{2020PhRvD.101f4011H}, namely
\begin{equation}\label{eq-s-12}
\begin{split}
  h=&{\mu_+}\left[A^+\cos(\omega_1t+\phi_1)+A^\times\sin\left(\omega_1t+\phi_1\right)\right]\\
  &+{\mu_-}\left[A^+\cos(\omega_2t+\phi_2)+A^\times\sin\left(\omega_2t+\phi_2\right)\right]\\
  =&\mu_s\Big[A^+\cos\left(\omega_\text{f}t+\phi_\text{f}\right)\cos\left(\omega_\text{b}t+\phi_\text{b}\right)\\
   &+A^\times\cos\left(\omega_\text{f}t+\phi_\text{f}-\frac{\pi}{2}\right)\cos\left(\omega_\text{b}t+\phi_\text{b}\right)\Big]+\\
  &\mu_d\left[A^+\cos\left(\omega_\text{f}t+\phi_\text{f}+\frac{\pi}{2}\right)\cos\left(\omega_\text{b}t+\phi_\text{b}-\frac{\pi}{2}\right)\right.\\
  &\left.+A^\times\cos\left(\omega_\text{f}t+\phi_\text{f}\right)\cos\left(\omega_\text{b}t+\phi_\text{b}-\frac{\pi}{2}\right)\right],
 \end{split}
\end{equation}
where $A^{+/\times}$ denote the polarization amplitudes, and $\mu_s=\mu_++\mu_-$ and $\mu_d=\mu_+-\mu_-$ are introduced for convenience. 

For a compact binary with component masses $m_1$ and $m_2$, the leading-order polarization amplitudes are approximately
\begin{equation}
\label{eq-as}
\begin{aligned}
A^+ &=
\frac{4\mathcal M_c}{d_L}
(\pi\mathcal M_c f)^{2/3}
F^+\frac{1+\cos^2\iota}{2},\\
A^\times &=
\frac{4\mathcal M_c}{d_L}
(\pi\mathcal M_c f)^{2/3}
F^\times\cos\iota,
\end{aligned}
\end{equation}
at the leading order, where $\mathcal M_c=(1+z_{\rm S})(m_1m_2)^{3/5}/(m_1+m_2)^{1/5}$ is the redshifted chirp mass,  $d_L$ is the luminosity distance, $\iota$ is the inclination angle and $F^{+/\times}$ are the antenna pattern functions \citep{2017arXiv171003794I}.

The quantities $\omega_1$ and $\omega_2$ represent the angular frequencies of the two lensed signals, while $\phi_1$ and $\phi_2$ are their initial phases. 
We further define
\begin{equation}
\label{eq-beat-def}
\begin{aligned}
\omega_\text{f} &= \frac{\omega_1+\omega_2}{2},
&\qquad
\phi_\text{f} &= \frac{\phi_1+\phi_2}{2},\\
\omega_\text{b} &= \frac{\omega_1-\omega_2}{2},
&\qquad
\phi_\text{b} &= \frac{\phi_1-\phi_2}{2}.
\end{aligned}
\end{equation}

During the early inspiral stage, the lensing time delay $\Delta t_{\rm lens}$ is sufficiently small that $\omega_\text{b}\ll\omega_\text{f}$. 
Under this condition, the superposition of the two lensed signals produces a beat pattern in the time domain characterized by the beat frequency $\omega_\text{b}$.

At leading post-Newtonian order, the GW frequency evolution of an inspiraling compact binary satisfies 
\begin{equation}\label{eq-t-eom}
  \frac{\ud\omega}{\ud t}=\frac{192}{5}\mathcal M_c^{5/3}\left( \frac{\omega}{2} \right)^{11/3}.
\end{equation}
Assuming $\omega_\text{b}\ll\omega_\text{f}$, one obtains
\begin{equation}\label{eq-ombt}
\omega_\text{b}\approx\frac{96}{5}\left(\frac{\omega_\text{f}}{2}\right)^{11/3}\mathcal M_c^{5/3}\Delta t_{\rm lens},
\end{equation}
where $\omega_\text{f}$ approximately corresponds to the instantaneous GW frequency. 
As the inspiral proceeds, $\omega_\text{f}$ increases continuously, implying that the ratio $\omega_\text{b}/\omega_\text{f}\propto\omega_\text{f}^{8/3}$ also grows with time. 
Consequently, the beat pattern gradually disappears at later stages of the evolution. 
After the merger signal carried by ray 1 passes the detector, its amplitude decreases rapidly, and the total strain becomes dominated by the second ray, such that $h\simeq h_2$. The remaining signal therefore behaves similarly to a standard unlensed waveform, apart from the additional magnification factor ${\mu_-}$.

\subsection{Condition for temporal overlap}
\label{sec-overlap-condition}

Compared with ground-based detectors, gravitational-wave signals in the LISA band can persist for months to years. In this regime, strongly lensed signals are not necessarily resolved as independent images. Their observational appearance is determined by the relative magnitude between the lensing-induced time delay and the intrinsic signal duration. The temporal-overlap condition can be written as
\begin{equation}
\Delta t_{\rm lens} < \Delta t_{\rm obs},
\end{equation}
which provides the condition for the formation of observable beat patterns. When this condition is satisfied, multiple lensed images overlap in time, leading to coherent superposition and producing characteristic beat patterns in the waveform. In contrast, when the time delay is comparable to or larger than the signal duration, the individual images are effectively separated and no observable beat pattern is generated.

To quantify the relevant timescale, we estimate the duration of the gravitational-wave signal from the leading-order post-Newtonian evolution of the binary system \citep{1963PhRv..131..435P}. The frequency evolution is given by
\begin{equation}
\frac{df}{dt} = \frac{96}{5}\pi^{8/3}
\left(\frac{G\mathcal M_c}{c^3}\right)^{5/3}
f^{11/3},
\end{equation}
which describes the chirp evolution during the inspiral phase. Integrating this equation yields the observable signal duration from a lower frequency cutoff \( f_{\rm low} \) to the innermost stable circular orbit (ISCO) frequency \( f_{\rm ISCO} \),
\begin{equation}
\Delta t_{\rm obs} =
\frac{5}{256}
\left(\frac{G\mathcal M_c}{c^3}\right)^{-5/3}
\pi^{-8/3}
\left(
f_{\rm low}^{-8/3}
-
f_{\rm ISCO}^{-8/3}
\right),
\end{equation}
whereas the ISCO frequency is given by
\begin{equation}
f_{\rm ISCO} =
\frac{c^3}{6^{3/2}\pi GM_{z,\rm tot}},
\end{equation}
where $M_{z,\rm tot}=(1+z_{\rm S})(m_1+m_2)$ is the redshifted total mass.

When \( f_{\rm low} > f_{\rm ISCO} \), the gravitational wave signal from the inspiral, merger, and ringdown phases is not fully contained within the LISA sensitivity band, and the source is only partially observable. In this case, the above expression provides an approximate estimate of the inspiral duration and is not strictly valid for the full waveform.

Since the beat modulation considered in this work is generated during the overlapping inspiral stage, we use the inspiral duration as the characteristic timescale for evaluating the temporal-overlap condition. The merger and ringdown phases are therefore not included in the duration estimate.

\subsection{Illustrative example}
\label{sec-example}

To illustrate the formation of beat patterns in the LISA band, we consider a representative strongly lensed binary black hole system generated using the \texttt{Triangle\_BBH} waveform model \citep{2026SCPMA..6949501D}. The source parameters are adopted from the simulated strongly lensed LISA source population presented by \citet{2025PhRvD.112l3512G}. The source is characterized by $z_{\rm S} = 5.00$, $m_1 = 9.13\times10^4\,M_\odot$, and $m_2 = 7.03\times10^4\,M_\odot$, with aligned dimensionless spin parameters $\chi_{1z} = 0.50$ and $\chi_{2z} = 0.30$.

The remaining source parameters are taken as $\phi_c = 5.32$ (coalescence phase), $\lambda_{\rm sky} = 5.04$ (ecliptic longitude), $\beta_{\rm sky} = 1.32$ (ecliptic latitude), and $\iota = 0.46$ (inclination angle between the orbital angular momentum and the line of sight). For the lensing configuration, we adopt amplitude magnification factors $\mu_+ = 2.24$ and $\mu_- = 2.00$ for the two lensed images, together with a lensing time delay of $\Delta t_{\rm lens} = 15.00~{\rm days}$.

Using the leading-order inspiral timescale estimate introduced above, the corresponding observable signal duration is found to be $\Delta t_{\rm obs} \simeq 148.89~{\rm days}$. This duration is substantially longer than the lensing time delay, implying that the two lensed signals overlap within the LISA observation band and can therefore produce observable beat patterns.

Figure~\ref{fig-int} shows a representative example of beat formation during the overlapping inspiral stage. Only the inspiral portion of the signal is displayed because the beat modulation is most coherent and visually prominent before merger. The waveforms are generated using the dominant $(2,2)$ mode of the \texttt{IMRPhenomD} model \citep{2016PhRvD..93d4006H,2016PhRvD..93d4007K}, neglecting higher-order modes and spin-precession effects. The top and middle panels correspond to the two lensed images, while the bottom panel shows their coherent superposition. A slowly varying amplitude envelope is clearly visible in the combined waveform, demonstrating the beat pattern generated by interference between the overlapping lensed signals.

\subsection{Frequency-domain beat modulation}
\label{sec-snr-g}

The beat phenomenon produced by the interference of two lensed gravitational-wave signals can also be described naturally in the frequency domain. Let $\tilde h_u(f)$ denote the frequency-domain waveform of the unlensed signal $h_u(t)$. For the first lensed image, the waveform amplitude is multiplied by the amplitude magnification factor $\mu_+$, giving \citep{2020PhRvD.101f4011H}
\begin{equation}
\tilde h_1(f)=\mu_+\tilde h_u(f).
\end{equation}

For the second image, the waveform experiences both an additional amplitude magnification factor $\mu_-$ and a lensing time delay $\Delta t_{\rm lens}$. Using the time-shift property of the Fourier transform, the corresponding frequency-domain waveform can be written as
\begin{equation}
\tilde h_2(f)=\mu_- e^{i2\pi f\Delta t_{\rm lens}}\tilde h_u(f).
\end{equation}

The total observed waveform is therefore given by the coherent superposition of the two lensed images,
\begin{equation}
\tilde h(f)=\tilde h_1(f)+\tilde h_2(f),
\end{equation}
which yields
\begin{equation}
\tilde h(f)=\left(\mu_+ + \mu_- e^{i2\pi f\Delta t_{\rm lens}}\right)\tilde h_u(f).
\end{equation}

Accordingly, the amplitude of the total frequency-domain waveform becomes
\begin{equation}
|\tilde h(f)|=\sqrt{\mu_+^2+\mu_-^2+2\mu_+\mu_-\cos(2\pi f\Delta t_{\rm lens})}\,|\tilde h_u(f)|.
\end{equation}

The oscillatory cosine term introduced by the relative time delay produces a characteristic modulation structure in the frequency domain, which corresponds to the beat pattern generated by the interference between the two lensed signals.

\section{Beat Signal Identification and Parameter Estimation}
\label{sec-methods}

\subsection{Mismatch for beat-pattern identification}
\label{sec-mismatch}

To quantify whether the lensing-induced beat pattern can be distinguished from a single lensed gravitational-wave image, we perform a matched-filtering analysis \citep{1994PhRvD..49.2658C,2023PhRvD.107j3023A}. In this work, the mismatch is not computed between the full two-image lensed waveform and a single-image waveform. Instead, we restrict the comparison to the time interval before the coalescence of the first image.

This choice is physically motivated. If the full two-image lensed waveform is used, the signal contains both the beat-modulated overlapping part and the later-arriving second merger peak. Comparing such a full waveform with a single lensed image would naturally produce a large mismatch, mainly because of the additional delayed merger signal rather than because of the beat modulation itself. This would overestimate the distinguishability of the beat pattern and may lead to a misleading conclusion that the beat signal is always very different from an ordinary lensed signal.

Therefore, to isolate the contribution of the beat modulation, we truncate both waveforms at the coalescence time of the first image. The reference waveform is taken to be the first lensed image alone, while the beat waveform is constructed from the coherent superposition of the two lensed images within the same time interval. In this way, the mismatch measures the morphological difference caused by the beat pattern during the overlapping inspiral stage.

For two waveforms $h_1$ and $h_2$, the mismatch is defined as \citep{1994PhRvD..49.2658C}
\begin{equation}
\epsilon(h_1,h_2)
=
1-
\max_{t_c,\phi_c}
\frac{\langle h_1|h_2\rangle}
{\sqrt{
\langle h_1|h_1\rangle
\langle h_2|h_2\rangle
}} .
\end{equation}

The noise-weighted inner product is given by
\begin{equation}
\langle h_1|h_2\rangle
=
4\,{\rm Re}
\int_{f_{\rm low}}^{f_{\rm cut}}
\frac{
\tilde h_1(f)\tilde h_2^*(f)
}{
S_n(f)
}\,df ,
\end{equation}
where $S_n(f)$ denotes the one-sided noise power spectral density of LISA, and $f_{\rm low}$ and $f_{\rm cut}$ are the lower and upper frequency bounds adopted in the calculation.

To quantify whether the beat modulation can be identified in the LISA data stream, we adopt the mismatch as a measure of waveform distinguishability. Following the standard criterion in gravitational-wave data analysis, two waveforms are considered distinguishable when the mismatch satisfies
\begin{equation}
\label{eq-mismatch-threshold}
\epsilon_{\rm beat}
\gtrsim
\frac{1}{\rho^2},
\end{equation}
where $\rho$ denotes the SNR of the observed signal \citep{1994PhRvD..49.2658C,1993PhRvD..47.2198F,1992PhRvD..46.5236F}.

This criterion reflects the fact that waveform differences smaller than the statistical uncertainty of the detector cannot be reliably resolved. Therefore, only when the mismatch exceeds the threshold $1/\rho^2$ can the beat modulation be regarded as observationally distinguishable from a single lensed gravitational-wave signal.

In this work, the mismatch is evaluated only during the overlapping inspiral stage before the coalescence of the first-arriving image,
\begin{equation}
\epsilon_{\rm beat}
=
\epsilon
\left[
h_{\rm beat}(t<t_{c,+}),
h_{\rm single}(t<t_{c,+})
\right],
\end{equation}
where $t_{c,+}$ is the coalescence time of the first image. This truncated comparison avoids the artificial mismatch contribution produced by the additional delayed merger peak of the second image, and therefore isolates the waveform deformation caused specifically by the beat modulation itself.

\subsection{Bayesian parameter estimation}
\label{sec-bayesian-parameter}

We further perform a Bayesian parameter-estimation analysis to test whether a beat-template model can recover the physical parameters encoded in a lensed beat signal. For the simulated data $d$, the posterior distribution of the sampled parameters is written as
\begin{equation}
p(\boldsymbol{\theta}|d)
=
\frac{
\mathcal L(d|\boldsymbol{\theta})p(\boldsymbol{\theta})
}{
\mathcal Z
},
\end{equation}
where $\mathcal L(d|\boldsymbol{\theta})$ is the likelihood, $p(\boldsymbol{\theta})$ is the prior, and $\mathcal Z$ is the Bayesian evidence,
\begin{equation}
\mathcal Z
=
\int
\mathcal L(d|\boldsymbol{\theta})p(\boldsymbol{\theta})
\,d\boldsymbol{\theta}.
\end{equation}
Here the sampled parameter vector is
\begin{equation}
\boldsymbol{\theta}
=
\{\mathcal M_c,q,\chi_{1z},\chi_{2z},t_c,\phi_c,
\log_{10}d_L,\mu_+,\mu_-,\Delta t_{\rm lens}\}.
\end{equation}
We first generate an unlensed massive black hole binary waveform and then construct a lensed beat signal by coherently superposing two images with amplitude magnification factors $\mu_+$ and $\mu_-$ and a lensing time delay $\Delta t_{\rm lens}$. The template used in the inference has the same two-image beat structure, so that the relative amplitude and phase modulation of the beat pattern directly constrain the delayed-image amplitude magnification and the lensing time delay. To simulate observational uncertainty, we add complex Gaussian noise in the frequency domain according to the LISA noise power spectral density. The noise PSDs are computed for the $A_2$, $E_2$, and $T_2$ time-delay interferometry (TDI) channels, where the subscript ``2'' denotes the second-generation TDI observables designed to suppress laser-frequency noise in the realistic LISA constellation with unequal and time-varying arm lengths \citep{2026SCPMA..6949501D,2023PhRvD.107h2001T,2025SSPMA..55w0410W}. Only the $A_2$ and $E_2$ channels are used in the final likelihood evaluation, while the $T_2$ channel is assigned zero weight in the inverse covariance matrix.

\section{Results}
\label{sec-results}

\subsection{Beat identification with mismatch}
\label{sec-results-mismatch}

We first apply the mismatch criterion in Eq. (\ref{eq-mismatch-threshold}) to determine when the lensing-induced modulation is distinguishable from an ordinary lensed waveform. The comparison is restricted to the overlapping inspiral stage, so that the measured mismatch is driven by the beat envelope rather than by the delayed merger peak of the second image. Systems satisfying $\epsilon_{\rm beat} \gtrsim 1/\rho^2$ are classified as beat-identifiable systems. This criterion allows us to map the region of source and lensing parameters in which the time delay, magnification ratio, signal duration, and SNR jointly produce an observable beat pattern in LISA.

\begin{figure*}[!t]
  \centering
  \IfFileExists{fig2.pdf}
    {\includegraphics[width=.48\textwidth]{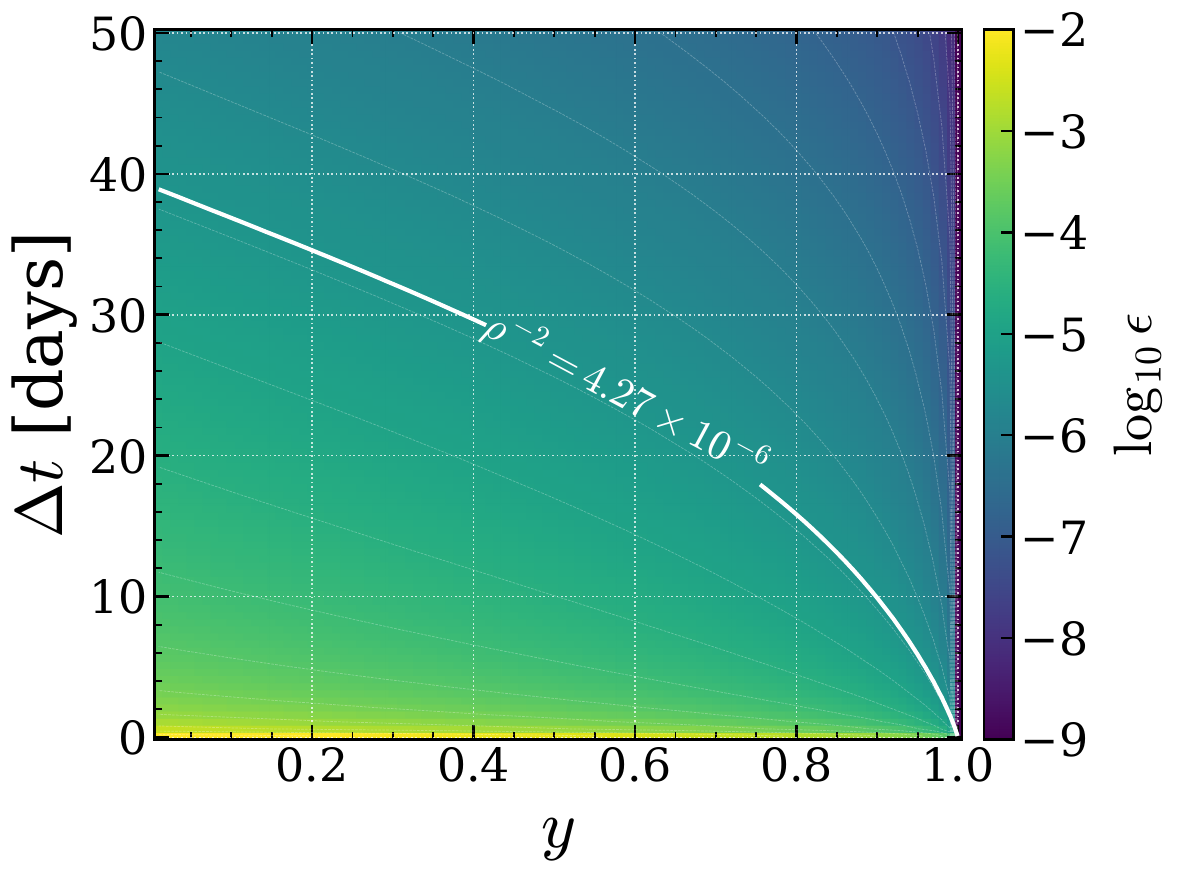}}
    {\fbox{\parbox[c][.32\textwidth][c]{.48\textwidth}{\centering fig2.pdf}}}
  \IfFileExists{fig3.pdf}
    {\includegraphics[width=.48\textwidth]{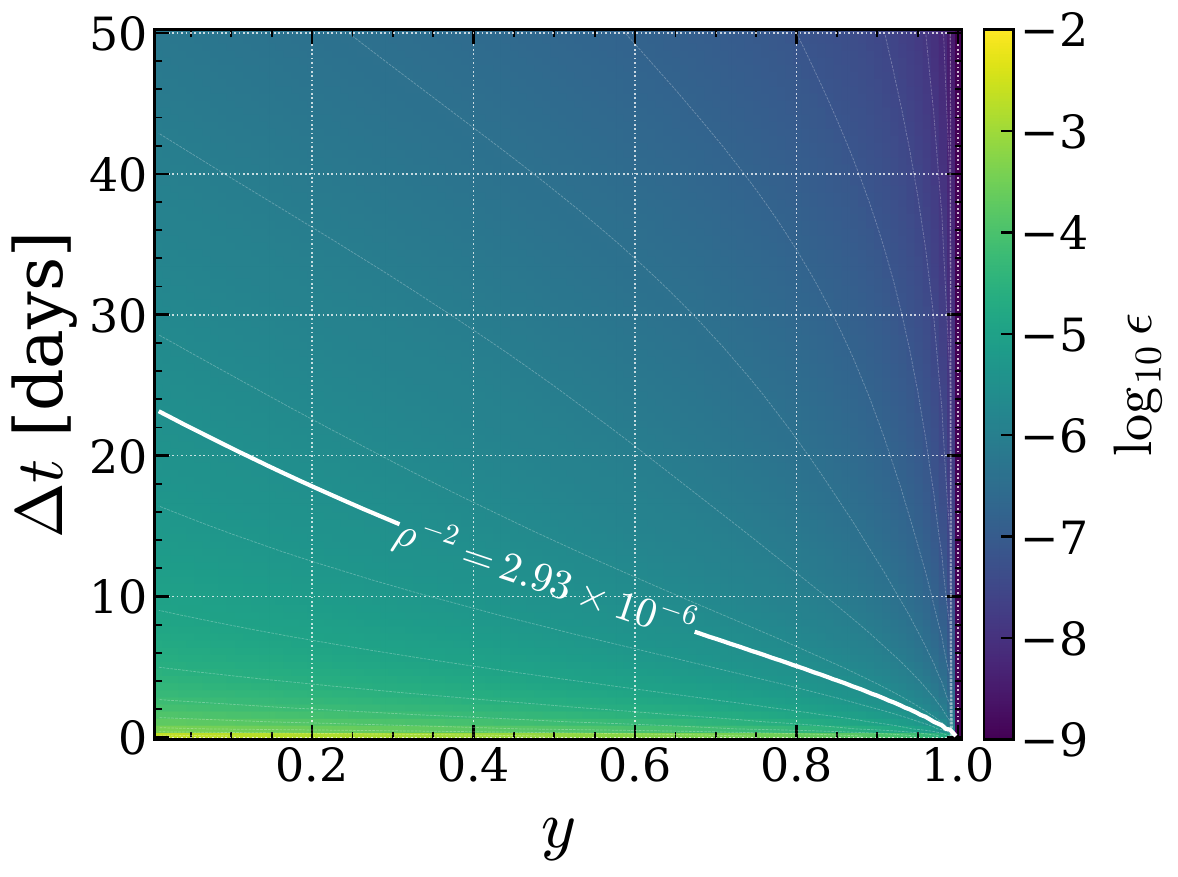}}
    {\fbox{\parbox[c][.32\textwidth][c]{.48\textwidth}{\centering fig3.pdf}}}
  \caption{Mismatch distributions for lensed gravitational-wave beat identification in the SIS model for two representative sources. The left and right panels correspond to two different massive black hole binary sources. The mismatch is shown as a function of the impact parameter $y$ and the lensing time delay $\Delta t_{\rm lens}$, denoted by $\Delta t$ in the plotted axis. The white solid lines denote the distinguishability threshold $\epsilon_{\rm beat}=1/\rho^2$; the regions below the lines correspond to beat-identifiable systems.}
  \label{fig-mismatch}
\end{figure*}

\begin{figure}[!t]
  \centering
  \IfFileExists{fig4.pdf}
    {\includegraphics[width=\columnwidth]{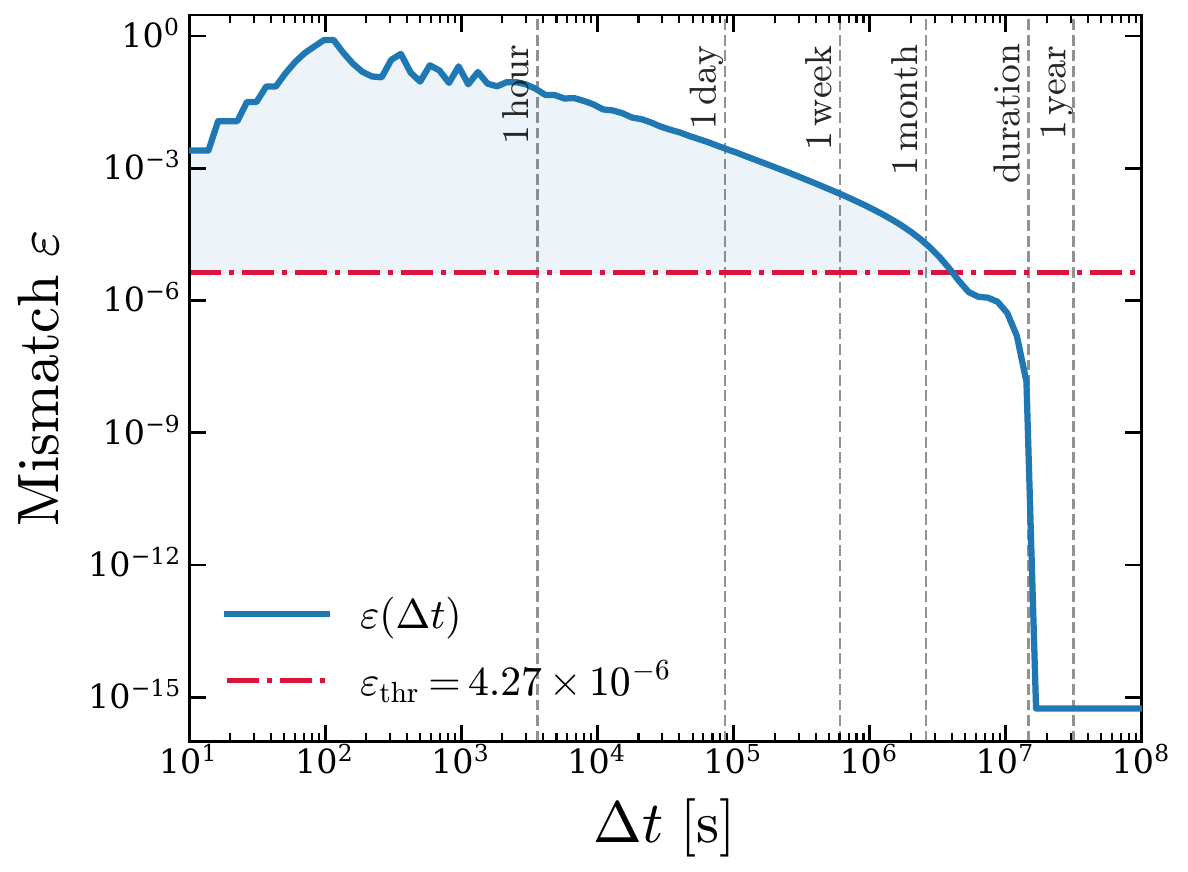}}
    {\fbox{\parbox[c][.60\columnwidth][c]{\columnwidth}{\centering mismatch\_vs\_time\_delay.pdf}}}
  \caption{Mismatch as a function of the lensing time delay for the same source parameters as the left panel of Fig.~\ref{fig-mismatch}. The blue curve shows $\epsilon(\Delta t_{\rm lens})$, while the red dash-dotted line denotes the distinguishability threshold $\epsilon_{\rm thr}=1/\rho^2$. The blue shaded region marks the range in which the beat modulation is identifiable. Vertical dashed lines indicate representative timescales.}
  \label{fig-mismatch-delay}
\end{figure}

Figure~\ref{fig-mismatch} shows the resulting mismatch distributions in the SIS model for two representative sources. The left panel uses a source with $z_{\rm S}=7.00$, $m_1=6.37\times10^4M_\odot$, $m_2=4.87\times10^4M_\odot$, $\chi_{1z}=0.22$, and $\chi_{2z}=-0.21$. The right panel uses a source with $z_{\rm S}=6.43$, $m_1=9.85\times10^4M_\odot$, $m_2=3.76\times10^4M_\odot$, $\chi_{1z}=-0.41$, and $\chi_{2z}=0.35$. Here we list only the key source parameters. The white solid lines denote the threshold $\epsilon_{\rm beat}=1/\rho^2$. The regions below the lines satisfy this criterion and correspond to beat-identifiable systems, whereas the regions above the lines have mismatches below the threshold.

The identifiable region is mainly controlled by the overlap time and the relative strength of the delayed image. A shorter time delay allows the two lensed signals to overlap more substantially during the inspiral stage, thereby enhancing the visibility of the beat modulation. The impact parameter $y$ determines the relative amplitudes of the two images. Using Eq.~(\ref{eq:sis_mag}), the amplitude ratio between the delayed and first-arriving images is $\frac{\mu_-}{\mu_+}=\left(\frac{1-y}{1+y}\right)^{1/2}.$ Therefore, as $y$ increases, the delayed image becomes progressively weaker relative to the first image. In this case, the delayed image contributes less to the total waveform, the interference effect is reduced, and the beat modulation becomes less significant. Consequently, the mismatch between the beat waveform and the single-image waveform decreases. Conversely, smaller $y$ gives two images with more comparable amplitudes, so the delayed image can modulate the total waveform more efficiently and produce a stronger beat envelope. Thus, beat patterns are easier to form and identify when the delayed image is sufficiently bright and the lensing time delay is sufficiently short. Detectable beat patterns are therefore expected only for a subset of strongly lensed LISA events with suitable lensing parameters and high enough SNR.

The role of the time delay is further illustrated in Fig.~\ref{fig-mismatch-delay}. For very small delays, the two images are almost fully overlapped, and increasing $\Delta t_{\rm lens}$ first makes their instantaneous phases and frequencies more different, causing the beat waveform to deviate more strongly from the single-image waveform. After this initial rise, the mismatch decreases steadily as the delay becomes longer, because the overlapping portion of the two signals becomes smaller and the coherent beat modulation is weakened. Once the delay approaches or exceeds the signal duration, the two images no longer overlap effectively in the analyzed inspiral interval. In this regime, the beat and reference waveforms become nearly indistinguishable in the truncated comparison, and the mismatch drops far below the threshold. Therefore, the beat pattern can be identified only over an intermediate range of time delays where the two images both overlap and remain sufficiently dephased to produce a measurable modulation.

\subsection{Expected fraction of beat events in LISA}
\label{sec-results-rate}

Using the HS-nod-SN (B+20) strongly lensed systems introduced in Sect.~\ref{sec-lensing-rate-model} as the parent population, we estimate the fraction of LISA strong-lensing events that can produce identifiable beat patterns. We sample 196 lensed gravitational-wave signals from the two-image detectable population. We first require the lensing time delay to be shorter than the observable signal duration, $\Delta t_{\rm lens}<\Delta t_{\rm obs}$, so that the two images can overlap in the LISA band. This temporal-overlap condition is satisfied by 92 events. We then compute the mismatch for these 92 overlapping systems and apply the distinguishability criterion $\epsilon_{\rm beat}\gtrsim1/\rho^2$. Among them, 14 events satisfy the mismatch criterion and are therefore classified as identifiable beat events.

\begin{table}[!t]
\caption{Selection of identifiable lensed beat events from the HS-nod-SN (B+20) two-image detectable sample.}
\label{tab-beat-rate}
\centering
\renewcommand{\arraystretch}{1.4}
\begin{tabular}{lcc}
\hline\hline
Selection step & Events & Fraction \\
\hline
Two-image lensed sample & 196 & 1.00 \\
Temporal overlap & 92 & 0.47 \\
Beat identifiable & 14 & 0.07 \\
\hline
\end{tabular}
\end{table}

Table~\ref{tab-beat-rate} summarizes this selection. The resulting probability of detecting an identifiable beat pattern among the two-image lensed events is therefore $0.07$. This value shows that lensed beat signals are a rare subset of LISA strong-lensing events: even when two lensed images are detectable, only systems with sufficient temporal overlap and sufficiently large beat-induced waveform deformation can be identified through the mismatch criterion.

\subsection{Parameter estimation from posterior distributions}
\label{sec-results-parameter}

\begin{figure*}[!t]
  \centering
  \IfFileExists{fig5.pdf}
    {\includegraphics[width=.95\textwidth]{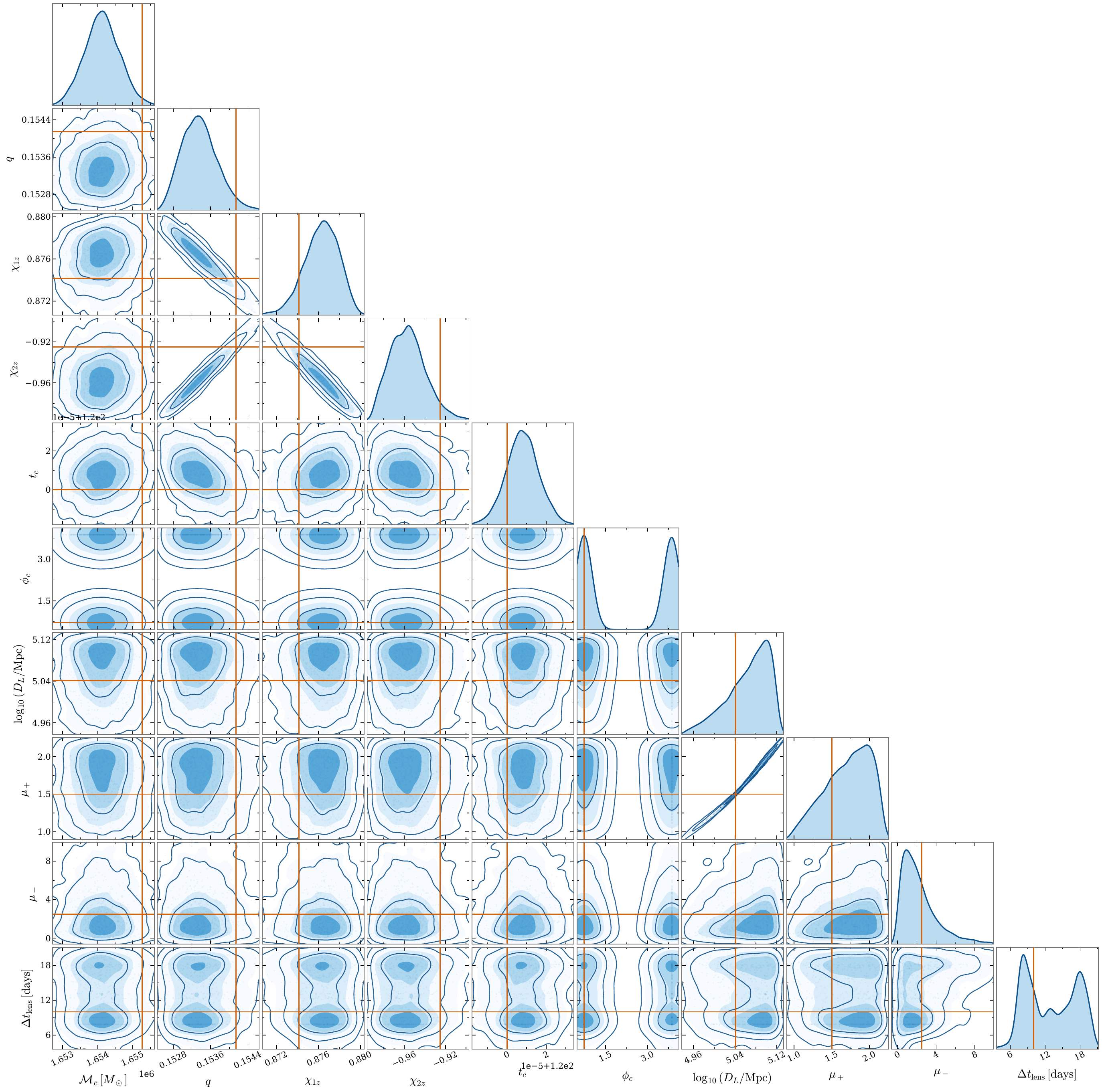}}
    {\fbox{\parbox[c][.60\textwidth][c]{.95\textwidth}{\centering blue\_kde\_corner\_reference\_style.pdf}}}
  \caption{Posterior corner plot for the Bayesian parameter-estimation analysis of a lensed beat signal using the beat template. The blue contours show the posterior distributions obtained with the $A_2$ and $E_2$ TDI channels, and the orange lines denote the injected parameter values.}
  \label{fig-corner}
\end{figure*}

Figure~\ref{fig-corner} shows the posterior distributions obtained for a representative identifiable beat event. The injected values are recovered within the posterior support, demonstrating that the beat template can successfully fit the beat signal. The key result is that the template recovers the lensing time delay $\Delta t_{\rm lens}$ and the delayed-image magnification $\mu_-$, which are the two parameters most directly responsible for the phase spacing and amplitude depth of the beat envelope. The recovery of $\mu_+$, $\mu_-$, and $\Delta t_{\rm lens}$ confirms that the beat modulation contains direct information about the relative amplitudes and arrival times of the lensed images. Therefore, identifiable beat signals can be used not only to recognize strongly lensed gravitational waves, but also to infer the key parameters of the lensing configuration from the beat pattern itself.

\section{Summary and Discussion}
\label{sec-discussion-conclusion}

In this section, we summarize the main results of this work and discuss the remaining issues that should be addressed in future applications to realistic LISA data.

First, we have studied strongly lensed gravitational-wave beat patterns produced by overlapping lensed images of massive black hole binaries in the LISA band. The long duration of LISA signals makes it possible for lensing time delays of days to months to be shorter than the observable inspiral duration, allowing coherent overlap and beat formation.

Second, we proposed an identification framework for lensed beat signals based on the mismatch between a beat waveform and a single-image waveform. The comparison is performed only before the coalescence of the first image, so that the criterion measures the beat modulation in the overlapping inspiral stage rather than the artificial difference caused by the delayed second merger peak. This truncated-signal strategy provides a cleaner way to decide whether the beat pattern itself is distinguishable.

Third, applying this criterion to the HS-nod-SN (B+20) LISA strong-lensing sample, we find that 92 out of 196 two-image lensed events satisfy the temporal-overlap condition $\Delta t_{\rm lens}<\Delta t_{\rm obs}$, while 14 events also satisfy the mismatch-identification criterion. This gives a beat-identification probability of $0.07$. The mismatch maps further show that beat signals are easier to form and identify when the lensing time delay is short and the delayed image has a sufficiently large magnification. In the SIS model, increasing the impact parameter $y$ weakens the delayed image relative to the first image, which reduces the interference strength and lowers the mismatch.

Finally, we built a Bayesian parameter-estimation template for beat signals and applied it to simulated LISA data using the $A_2$ and $E_2$ TDI channels. The posterior distributions show that the beat template can recover the lensing time delay and the magnification of the delayed image, together with the other sampled source and lensing parameters. This demonstrates that an identifiable beat pattern can be used not only as a lensing signature, but also as a carrier of quantitative information about the relative arrival time and amplitude of the two images.

The mismatch and posterior analyses address complementary aspects of lensed beat signals. The mismatch quantifies whether the beat modulation is distinguishable from an ordinary lensed waveform, while the posterior distributions illustrate how the beat structure constrains source and lensing parameters. The mismatch criterion used here is a detectability proxy rather than a full Bayesian model-selection calculation, but it provides a useful first criterion for selecting systems in which the beat modulation should be modeled explicitly.

In this work, the two overlapping waveforms are assumed to be two lensed images of the same gravitational-wave source. In real LISA data, however, unrelated massive black hole binary signals may also appear in the same time window. Such an accidental overlap would not obey the same source-parameter consistency and lensing relations as a true pair of lensed images, but it could still produce apparent amplitude modulation in part of the data stream. Therefore, a practical search should combine the beat-identification framework with additional consistency tests, including common intrinsic parameters, compatible sky localization, coherent phase evolution, and lensing relations among $\mu_+$, $\mu_-$, and $\Delta t_{\rm lens}$. This source-consistency problem is an important step toward applying the present framework to a realistic LISA catalogue.

\begin{acknowledgments}
This work was supported by National Key R$\&$D Program of China (No. 2024YFC2207400).
\end{acknowledgments}

\bibliographystyle{apsrev4-2-five-authors}
\bibliography{GW}
\end{document}